\documentclass[aps,twocolumn,pra,tightenlines,floatfix,superscriptaddress]{revtex4-2}

\usepackage[breaklinks=true]{hyperref}
\usepackage{graphicx}
\usepackage[english]{babel}
\usepackage{amsmath}
\usepackage{amssymb}
\usepackage{times}

\begin{document}

\newcommand{\D}{\mathrm{D}}
\newcommand{\p}{\partial}
\renewcommand{\d}{\mathrm{d}}
\newcommand{\Ek}{E_\mathbf{k}}
\newcommand{\sumk}{\sum_\mathbf{k}}

\title{Ground states of atomic Fermi gases in a two-dimensional
  optical lattice with and without population imbalance}

\author{Lin Sun}
\affiliation{Department of Physics and Zhejiang Institute of Modern Physics, Zhejiang University, Hangzhou, Zhejiang 310027, China}

\author{Qijin Chen}
\email[Corresponding author: ]{qchen@uchicago.edu}
 \affiliation{Hefei National Research Center for Physical Sciences at the Microscale and School of Physical Sciences, University of Science and Technology of China,  Hefei, Anhui 230026, China}
\affiliation{Shanghai Research Center for Quantum Science and CAS Center for Excellence in Quantum Information and Quantum Physics, University of
  Science and Technology of China, Shanghai 201315, China}
%\affiliation{Hefei National Laboratory, Hefei 230088, China}
%\affiliation{Shanghai Branch, Hefei National Laboratory, Shanghai 201315, China}
\affiliation{Hefei National Laboratory, University of
  Science and Technology of China, Hefei 230088, China}

\date{\today}

\begin{abstract}
  We study the ground state phase diagram of population balanced and
  imbalanced ultracold atomic Fermi gases with a short range
  attractive interaction throughout the crossover from BCS to
  Bose-Einstein condensation (BEC), in a two-dimensional optical
  lattice (2DOL) comprised of two lattice and one continuum
  dimensions.  We find that the mixing of lattice and continuum
  dimensions, together with population imbalance, has an extraordinary
  effect on pairing and the superfluidity of atomic Fermi gases. In
  the balanced case, the superfluid ground state prevails the majority
  of the phase space. However, for relatively small lattice hopping
  integral $t$ and large lattice constant $d$, a pair density wave
  (PDW) emerges unexpectedly at intermediate coupling strength, and
  the nature of the in-plane and overall pairing changes from
  particle-like to hole-like in the BCS and unitary regimes,
  associated with an abnormal increase in the Fermi volume with the
  pairing strength.  In the imbalanced case, the stable polarized
  superfluid phase shrinks to only a small portion of the entire phase
  space spanned by $t$, $d$, imbalance $p$ and interaction strength
  $U$, mainly in the bosonic regime of low $p$, moderately strong
  pairing, and relatively large $t$ and small $d$.  Due to the Pauli
  exclusion between paired and excessive fermions within the confined
  momentum space, a PDW phase emerges and the overall pairing evolves
  from particle-like into hole-like, as the pairing strength grows
  stronger in the BEC regime. In both cases, the ground state property
  is largely governed by the Fermi surface topology. These findings
  are very different from the cases of pure 3D continuum, 3D lattice
  or 1DOL.
\end{abstract}

\maketitle

\section{Introduction}
%Introduction
  %Background

Ultracold Fermi gases provide an ideal platform for investigating the
pairing and superfluid physics over the past decades, primarily owing
to the high tunability of multiple parameters
\cite{chen2005PR,bloch2008RMP}.  Using a Feshbach resonance
\cite{chin2010RMP}, one can tune the effective pairing strength from
the weak coupling BCS limit all the way through to the strong pairing
Bose-Einstein condensation (BEC) limit.  There have been a great
number of experimental and theoretical studies on ultracold Fermi
gases in recent years, with many tunable parameters which have been
made accessible experimentally, including pairing interaction strength
\cite{chen2005PR}, population imbalance
\cite{zwierlein2006S,partridge2006S,FGLW05,PWY05,YD05,chen2006PRA,SM06,Haque2006,radzihovsky2010RoPP},
and dimensionality \cite{martiyanov2010PRL,wu2015PRL,zhang2017SR}.  In
particular, ultracold Fermi gases in an optical lattice exhibit rich
physics due to the tunable geometry
\cite{bloch2005NP,kohl2005PRL,schneider2008S}.  As is well known,
population imbalance suppresses or destroys superfluidity in
three-dimensional (3D) homogeneous systems
\cite{chen2006PRA,chien2006PRL}.  For example, superfluidity at zero
temperature is completely destroyed at unitarity and in the BCS
regime, whereas stable polarized superfluid (pSF) with a finite
imbalance $p$ exists only in the BEC regime \cite{chien2006PRL}.
Meanwhile, in the absence of population imbalance in a 3D lattice, one
finds the superfluid transition temperature
$T_\text{c} \propto -t^2/U$ in the BEC regime, due to virtual pair
unbinding in the pair hopping process \cite{NSR,chen1999PRB}, which
makes it hard to reach the superfluid phase in the BEC regime. (Here
$t$ is the lattice hopping integral, and $U<0$ is the onsite
attractive interaction). While the superfluid transition for both
population balanced and imbalanced Fermi gases have been realized
experimentally in the 3D continuum case (often in a trap), it has not
been realized even for the balanced case in 3D lattices. However,
superfluidity, long-range or Berezinskii-Kosterlitz-Thouless
(BKT)-like \cite{Berezinskii,*Kosterlitz}, as well as pairing
phenomena have been explored experimentally in 2D and 1D optical
lattices
\cite{Martin12PRL,liao2010N,pagano2014NP,revelle2016PRL,ThomasPRA94,Sobirey2021}
or quasi-2D traps
\cite{Jochim1,Jochim2,PhysRevLett.117.093601,Refth0,PhysRevLett.120.060402}. Common
to these experiments is the the presence of one or two continuum
dimensions. Until further breakthrough is made in cooling techniques,
the presence of continuum dimensions seems to be crucial for the
superfluid phase to be accessible experimentally so far in low
dimensional optical lattices (and quasi-2D traps) besides the 3D
continuum. We note, however, that these optical lattice experiments
have mostly been restricted to the small $t$ limit such that the
coupling between different pancakes (2D planes) or cigar-shaped tubes
(1D lines) is negligible. Therefore, a systematic investigation of the
vast unexplored parameter space of the low dimensional optical
lattices is important in order to uncover possible exotic and
interesting new quantum phenomena.

In the presence of population imbalance, an open Fermi surface of
Fermi gases in a one-dimensional optical lattice (1DOL), caused by
large $d$ and/or small $t$, often leads to destruction of the
superfluid ground state in the BEC regime \cite{wang2020PRA2}.  Our
recent study on pairing and superfluidity of atomic Fermi gases in a
two-dimensional optical lattice (2DOL), which is comprised of two
lattice and one continuum dimensions, reveals that for relatively
large $d$ and small $t$, a pair density wave (PDW) ground state
emerges in the regime of intermediate pairing strength, and the nature
of the in-plane and overall pairing changes from particle-like to
hole-like in the unitary and BCS regimes, with an unexpected
nonmonotonic dependence of the chemical potential on the pairing
strength \cite{sun2021AdP}.

In this paper, we focus on the ground state superfluid behavior of
atomic Fermi gases in 2DOL, under the effects of lattice-continuum
mixing, population imbalance and its interplay with the lattice
parameters.  We first investigate the evolution of the Fermi surface
as a function of hopping integral $t$ and lattice constant $d$, and
then calculate the zero $T$ superfluid phase diagram using the
BCS-Leggett mean-field equations \cite{leggett1980}, but supplemented
with various stability conditions, including those derived from
finite-temperature formalism \cite{chen2006PRA}. We explore the
superfluid phase diagrams in various phase planes, as a function of
lattice constant, hopping integral and interaction strength for
population balanced cases and also of polarization for population
imbalanced cases.

We find that in the population balanced case, while the phase diagram
at zero $T$ is dominated by the superfluid phase, a PDW ground state
may emerge at intermediate pairing strength, for relatively small $t$
and large $d$, and the nature of the in-plane and overall pairing
changes from particle-like to hole-like in the BCS and unitary
regimes. This is associated with an open Fermi surface, where the
effective number density in the lattice dimensions can go above half
filling. The PDW state originates from strong inter-pair repulsive
interactions and relatively large pair size at intermediate pairing
strength, which is also found in dipolar Fermi gases within the
pairing fluctuation theory \cite{che2016PRA}.

In the population imbalanced case, due to the constraint of various
stability conditions, stable superfluid ground states are found to
exist only in a small portion of the multi-dimensional phase space,
spanned by the parameters $t$, $d$, $p$ and $U$, mainly in the low $p$
and bosonic regime of intermediate pairing strength, and for
relatively large $t$ and small $d$.  As the pairing interaction
becomes stronger in the BEC regime, the nature of the overall pairing
of a polarized Fermi gas in 2DOL evolves from particle-like into
hole-like.  As manifested in the momentum distribution of the paired
fermions and excessive majority fermions, there is a strong Pauli
exclusion between them for small $t$ and large $d$. Therefore,
decreasing $t$ and increasing $d$ and $p$ help to extend the hole-like
pairing regime toward weaker coupling.  These results are very
different from their counterpart in pure 3D continuum, 3D lattices and
1DOL.

We mention that the values of $t$ and $d$ for which one finds
hole-like pairing in the weaker coupling regime in the balanced case
and in the stronger coupling regime in the imbalanced case do not
overlap. This can be understood as the balanced case and the
$p\rightarrow 0^+$ case are not continuously connected at $T=0$.

\section{Theoretical formalism}
\subsection{General theory}
%Theoretical formalism
%Mean-field theory

Here we consider a two-component ultracold Fermi gas with a
short-range pairing interaction, $V_\mathbf{k,k'} = U<0$, in 2DOL.
The dispersion of noninteracting atoms without population imbalance is
given by
$\xi_\mathbf{k}=\epsilon_\mathbf{k}-\mu \equiv{k_z}^2/2m + 2t[2-\cos
(k_xd)-\cos (k_yd)]-\mu$, where $k_z$ is the momentum in the $z$
direction in the continuum dimension, $k_x$ and $k_y$ are the momenta
in the lattice plane, $t$ and $d$ are the hopping integral and lattice
constant in the $xy$ plane, respectively, and $\mu$ is the chemical
potential.  Following our recent works
\cite{wang2020PRA,wang2020PRA2,zhang2017SR,Zhang2020SCPMA}, we take
$t$ to be physically accessible, under the constraint $2mtd^2 < 1$ in
our calculation.  And the critical coupling for forming a two-body
bound state of zero binding energy is given by
$U_\text{c}=-1/\sum_{\textbf{k}}1/2\epsilon_{\textbf{k}} =
-0.16072\sqrt{2m}/\sqrt{t}d^2$.  Here and throughout we take the
natural units and set $\hbar=k_\text{B}=1$.

At zero temperature, the mean-field BCS-Leggett ground state follows
the gap and number equations \cite{leggett1980}
\begin{eqnarray}
  0&=&\frac{1}{U}+\sum_{\textbf{k}}\frac{1}{2E_{\textbf{k}}}\,,
  \label{eq:gap}\\
  n&=&\sum_{\textbf{k}}\Big(1-\frac{\xi_{\textbf{k}}}{E_{\textbf{k}}}\Big)\,, 
\label{eq:neq}
\end{eqnarray}
where $E_{\textbf{k}}=\sqrt{\xi_{\textbf{k}}^{2}+\Delta^{2}}$ is the
Bogoliubov quasiparticle dispersion, with an energy gap $\Delta$.

To make sure the mean-field solution is stable, we impose the
requirement that the dispersion of the Cooper pairs be
non-negative, both in the lattice plane and along the $z$ direction. To
this end, we extract the inverse pair mass (tensor) using the
fluctuating pair propagator, as given in the pairing fluctuation
theory which was previously developed for the pseudogap physics in the
cuprates \cite{chen1998PRL} and extended to address the BCS-BEC
crossover in ultracold atomic Fermi gases \cite{chen2005PR}. In
particular, we mention that, compared to rival $T$-matrix
approximations for the pairing physics, the pair dispersion as
extracted from this theory is gapless below $T_\text{c}$, fully compatible
with the mean-field gap equation. Here the pairing $T$ matrix
is given by
$t_\text{pg}(Q) = U/[1+U\chi(Q)]$, with the pair susceptibility
$\chi(Q)=\sum_K G_0(Q-K)G(K)$, the bare Green's function
$G_0(K) = (\omega-\xi_{\mathbf{k}})^{-1}$, and the full Green's
function
$G(K)=\frac{u_{\textbf{k}}^{2}}{\omega-E_{\textbf{k}}}+\frac{v_{\textbf{k}}^{2}}{\omega+E_{\textbf{k}}}$, where $u_{\textbf{k}}^{2}=(1+\xi_{\textbf{k}}/E_{\textbf{k}})/2$ 
and $v_{\textbf{k}}^{2}=(1-\xi_{\textbf{k}}/E_{\textbf{k}})/2$ are the BCS coherence factors, and 
$K\equiv (\omega, \mathbf{k})$, $Q\equiv (\Omega, \mathbf{q})$ are four momenta.

The inverse $T$-matrix $t_\text{pg}^{-1}(Q)$ can be expanded for small
$Q$, given by
$t_\text{pg}^{-1}(\Omega,\textbf{q})\approx
a_1\Omega^{2}+a_{0}(\Omega-\Omega_{\textbf{q}}+\mu_{p})$, with
$\Omega_{\textbf{q}}=B{q_z}^{2} +
2t_{B}[2-\cos(q_{x}d)-\cos(q_{y}d)]$, and $\mu_p=0$ in the superfluid
phase.  Then we extract $B=1/2M$, with $M$ being the effective pair
mass in the $z$ direction, and $t_{B}$ the effective pair hopping
integral in the $xy$ plane.  The sign of $a_0$ determines whether the
fermion pairs are particle-like or hole-like, with positive $a_0$ for
particle-like pairing and negative $a_0$ for hole-like pairing. For
example, in a 3D lattice, in general one finds $a_0 > 0$ for fermion
density below half filling, $a_0=0$ at half filling due to
particle-hole symmetry, and $a_0<0$ above half filling. The sign of
$a_0$ is controlled by the average of the inverse band mass
\footnote{For the latter case, one can perform a particle-hole
  transfermation so that it becomes $a_0>0$ and below half filling for
  holes.}. While one could perform a particle-hole transformation for
a pure lattice case, it does not seem to be feasible in our case
since both lattice and continuum dimensions are present. The
expressions for the coefficients $a_{1}$, $a_{0}$, $B$ and $t_{B}$ can
be readily derived during the Taylor expansion. In this way, using the
solution for ($\mu$, $\Delta$) from
Eqs. (\ref{eq:gap})-(\ref{eq:neq}), we can extract the pair dispersion
$\tilde{\Omega}_{\textbf{q}}=(\sqrt{a_0^{2}+4a_1a_0\Omega_{\textbf{q}}}-a_0)/2a_1$.
The non-negativeness of the pair dispersion implies that the pairing
correlation length (squared) $\xi^2 = a_0 B$ and
$\xi_{xy}^2 = a_0t_\text{B}d^2$ must be positive.

For the population imbalanced case, the spin polarization is defined
via $p=(n_\uparrow-n_\downarrow)/(n_\uparrow+n_\downarrow)$, where
spin index $\sigma=\uparrow,\downarrow$ refers to the majority and
minority components, respectively.  Then the dispersion of
noninteracting atoms is modified as
$\xi_{\mathbf{k}\sigma}=\epsilon_\mathbf{k}-\mu_\sigma$
% ={k_z}^2/2m + 2t[2-\cos (k_xd)-\cos (k_yd)]-\mu_\sigma$
, with $\mu_\sigma$ the chemical potential for spin $\sigma$.

Now the bare and full Green's functions are given by
\begin{eqnarray}
G_{0\sigma}(K) &=& 
\frac{1}{\omega-\xi_{\mathbf{k}\sigma}}\,,\quad\text{and}\nonumber\\
G_{\sigma}(K)&=&\frac{u_{\textbf{k}}^{2}}{\omega-E_{\textbf{k}\sigma}}+\frac{v_{\textbf{k}}^{2}}{\omega+E_{\textbf{k}\bar{\sigma}}},\nonumber
\end{eqnarray}
respectively, where $\bar{\sigma}$ is the opposite spin of $\sigma$, $E_{\textbf{k}\uparrow}=E_{\textbf{k}}-h$, and
$E_{\textbf{k}\downarrow}=E_{\textbf{k}}+h$, with  $\mu=(\mu_{\uparrow}+\mu_{\downarrow})/2$, and 
$h=(\mu_{\uparrow}-\mu_{\downarrow})/2$. Thus $E_{\textbf{k}\uparrow}$ becomes gapless, as it should, in order to accommodate the excessive majority fermions [See Eq.~(\ref{eq:neqb}) below]. These gapless fermions will contribute in both the gap and number equations.

Following the BCS self-consistency condition and the number constraint, we arrive at the gap and number equations at zero $T$ in the presence of population imbalance: 
\begin{eqnarray}
  0&=&\frac{1}{U}+\sum_{\textbf{k}}\frac{\Theta(E_{\textbf{k}\uparrow})}{2E_{\textbf{k}}}
  % \sum_{\textbf{k}}\Big[\frac{1-\Theta(-E_{\textbf{k}\uparrow})}{2E_{\textbf{k}}}\Big]
       \,,
  \label{eq:pgap}\\
  n&=&\sum_{\mathbf{k}}\Big[\Big(1-\frac{\xi_{\mathbf{k}}}{E_{\mathbf{k}}}\Big)+\Theta(-E_{\textbf{k}\uparrow})\frac{\xi_{\mathbf{k}}}{E_{\mathbf{k}}}\Big]\,,
  \label{eq:neqa}\\
  p n&=&\sum_{\mathbf{k}}\Theta(-E_{\textbf{k}\uparrow})\,,
  \label{eq:neqb}
\end{eqnarray}
where $\Theta(x)$ is the Heaviside step function, and 
%$E_{\textbf{k}\uparrow}=E_{\textbf{k}}-h$, 
%$E_{\textbf{k}\downarrow}=E_{\textbf{k}}+h$, and 
%$E_{\textbf{k}}=\sqrt{\xi_{\textbf{k}}^{2}+\Delta^{2}}$, 
%$\xi_{\textbf{k}}=\epsilon_{\textbf{k}}-\mu$, 
%$\mu=(\mu_{\uparrow}+\mu_{\downarrow})/2$, 
%$h=(\mu_{\uparrow}-\mu_{\downarrow})/2$.
%And we have the total fermion number density
$n = n_\uparrow + n_\downarrow$ and
$\delta n= n_{\uparrow}-n_{\downarrow}=pn$ are the total  and
the  difference of fermion densities, respectively.

In the imbalanced case, the pair susceptibility  is modified as
$\chi(Q) = \sum_{K,\sigma}G_{0\sigma}(Q-K)G_{\bar{\sigma}}(K)/2$,
which is consistent with the BCS self-consistency condition so that
the pair dispersion remains gapless at $q=0$. Then we follow the same
procedure as in the balanced case, and extract the inverse pair mass
tensor along with coefficients $a_0$ and $a_1$ via the Taylor
expansion of the inverse $T$ matrix, $t_\text{pg}^{-1}(Q)$.

Equations (\ref{eq:pgap})-(\ref{eq:neqb}) form a closed set of
self-consistent equations, and can be used to solve for ($\mu$, $h$,
$\Delta$) as a function of ($U$, $t$, $d$, $p$), which is then further constrained by various stability conditions.

% Thermodynamics
\subsection{Stability analysis}

As shown in the 3D continuum and 1DOL cases, in the presence of population
imbalance, not all solutions of Eqs. (\ref{eq:pgap})-(\ref{eq:neqb})
are stable \cite{chien2006PRL,chen2006PRA,Chen2020cpl}.

Following the stability analysis of
Refs.~\cite{chien2006PRL,chen2006PRA}, the stability condition for the
superfluid phase requires that for fixed $\mu$ and $h$, the solution
for the excitation gap $\Delta$ is a minimum of the thermodynamic
potential $\Omega_\text{S}$, which is demonstrated to be equivalent to
the positive definiteness of the %particle number susceptibility
generalized compressibility matrix \cite{chen2006PRA,pao2006PRB}.
%
% And the thermodynamic potential $\Omega_\text{S}$ are consists of the fermionic ($\Omega_\text{F}$) and bosonic ($\Omega_\text{B}$) contributions, 
% \begin{eqnarray}
%   \Omega_\text{S}&=&\Omega_\text{F}+\Omega_\text{B}\,,\nonumber\\
%   \Omega_\text{F}&=&-\frac{\Delta^{2}}{U}+\sum_{\mathbf{k}}(\xi_{\mathbf{k}} -E_{\mathbf{k}})-T\sum_{\mathbf{k},\sigma}\ln (1 + e^{-E_{\mathbf{k}\sigma}/T})\,,\nonumber\\
%   \Omega_\text{B}&=& T\sum_{\mathbf{q}}\ln(1-e^{-\tilde{\Omega}_{\mathbf{q}}/T})\,.
% \end{eqnarray}
%
Thus we have 
\begin{equation}
  \frac{\partial^{2}\Omega_\text{S}}{\partial\Delta^{2}}=\sum_{\mathbf{k}}\frac{\Delta^2}{E_{\mathbf{k}}^2}\Big[\frac{\Theta(E_{\textbf{k}\uparrow})}{E_{\mathbf{k}}}
  -\delta(E_{\textbf{k}\uparrow})
  \Big]>0\,,
  \label{eq:stab}
\end{equation}
where  $\delta(x)$ is the  delta function.% $\delta(x)=\mathrm{d} \Theta(x)/ \mathrm{d}x$. 

\begin{figure}
%  \centerline{\includegraphics[clip,width=3.4in]{PairDispersion_2.pdf}}
  \centerline{\includegraphics[clip,width=3.4in]{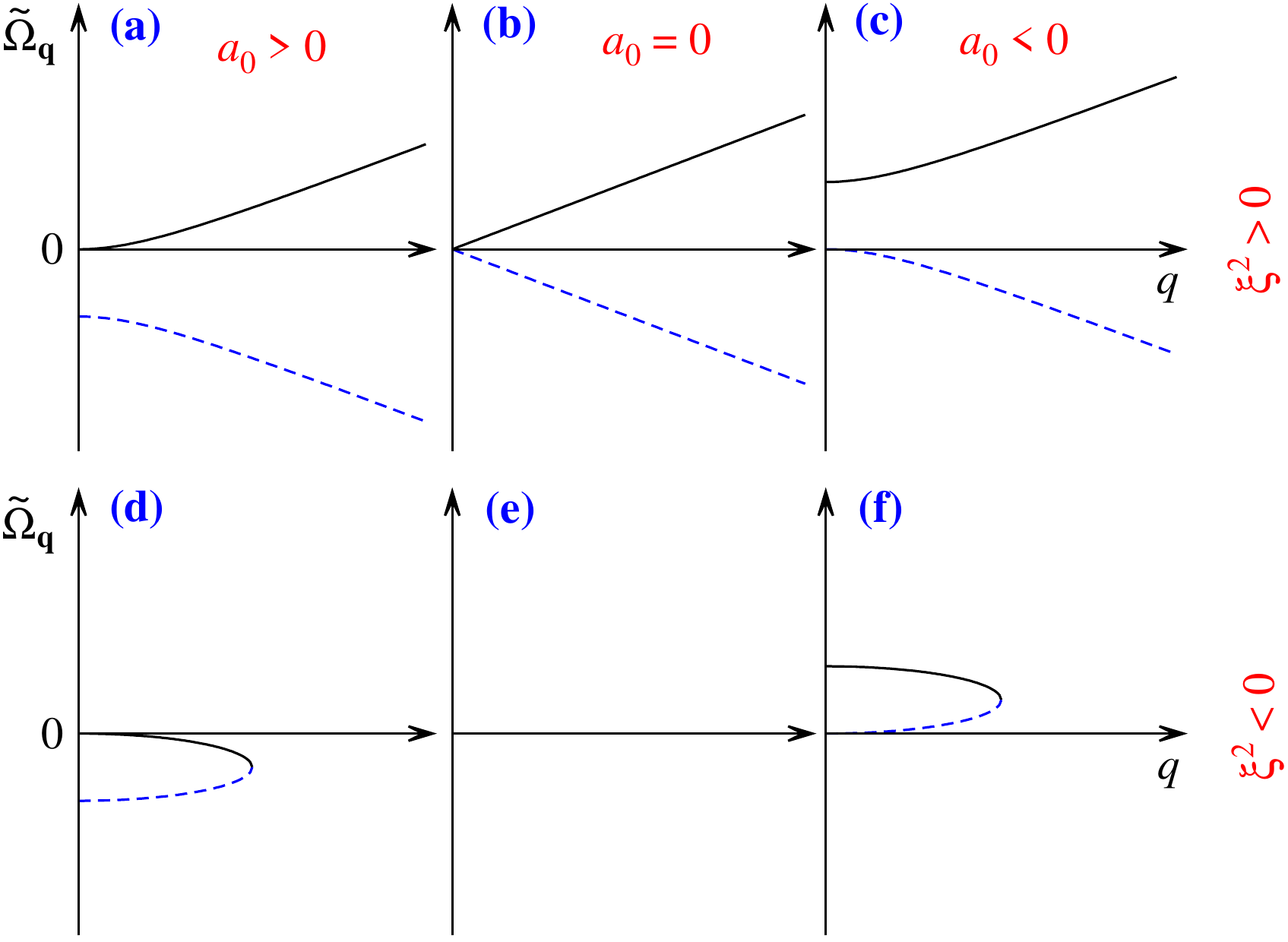}}
  \caption{Qualitative behavior of the pair dispersion $\tilde{\Omega}_\mathbf{q}$ for different signs of $a_0$ and $\xi^2$. For illustration purpose, a simple isotropic quadratic $\Omega_\mathbf{q}=\xi^2q^2/a_0$ is used. The three columns are for  $a_0>0$, $a_0=0$ and $a_0<0$ from left to right, and the top and bottom rows are $\xi^2>0$ and $\xi^2<0$, respectively. The black solid curves in the top row represent propagating modes.}
  \label{fig:Fig0}
\end{figure}

\begin{figure*}
  % \centerline{\includegraphics[clip,height=3.0in]{FS_t0.05d1&2&3&4_d3t0.01&0.04&0.07&0.1_4col2row.pdf}}
%  \centerline{\includegraphics[clip,width=6.in]{Fig1.pdf}}
  \centerline{\includegraphics[clip,width=6.in]{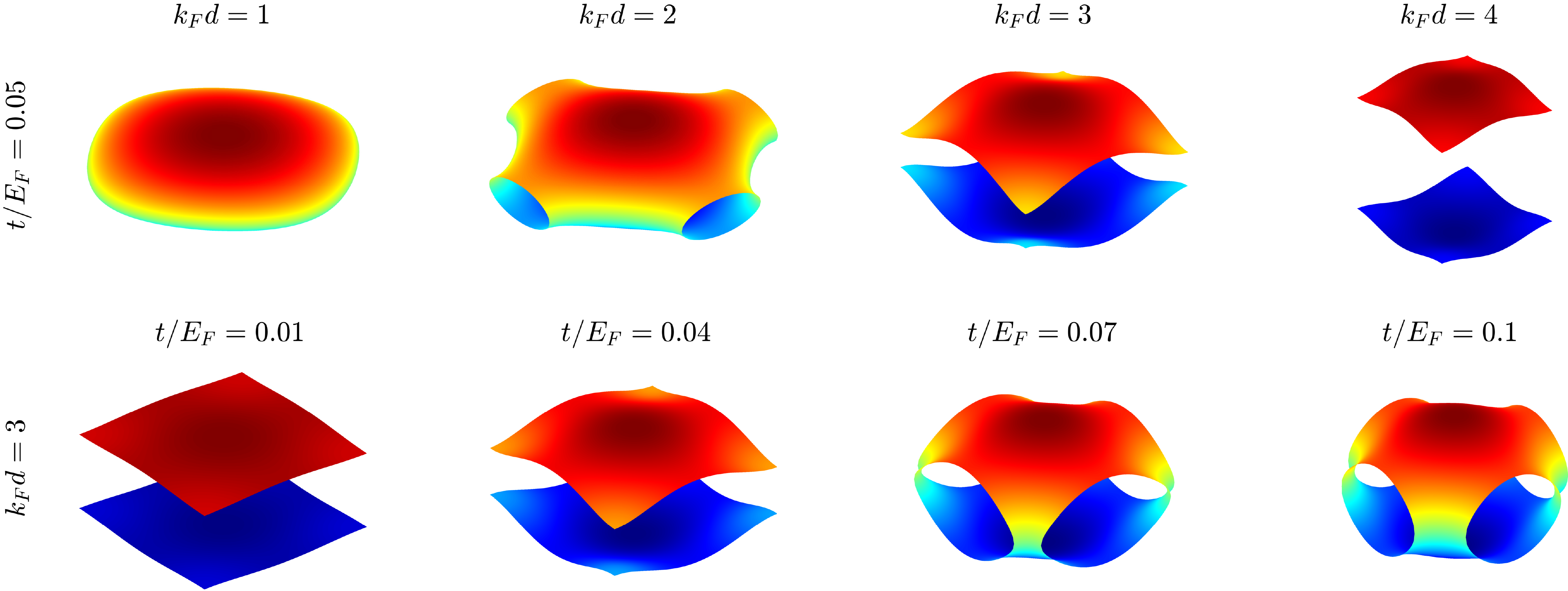}}
  \caption{Evolution of the Fermi surface of atomic Fermi gases in
    2DOL, for fixed $t/E_\text{F}=0.05$ (top
    row) with $k_\text{F}d=1$, $2$, $3$ and $4$, and fixed
    $k_\text{F}d=3$ (bottom row) with $t/E_\text{F}=0.01$, $0.04$,
    $0.07$ and $0.1$, from left to right. }
  \label{fig:Fig1}
\end{figure*}

In addition, the positivity of the pair dispersion in the entire
momentum space imposes another strong stability condition. Illustrated
in Fig.~\ref{fig:Fig0} are the qualitative behaviors of the pair
dispersion, for different signs of $a_0$ and $\xi^2$. For illustration
purpose, a simple isotropic quadratic dispersion is assumed. In
general, there are two branches of the dispersion, from the inverse
$T$-matrix expansion up to the $\Omega^2$ order. The positive branch
represents a propagating mode, while the negative branch represents a
hole-like mode which contributes to quantum fluctuations. The case of
$a_0>0$ and $\xi^2>0$ (Fig.~\ref{fig:Fig0}(a)) corresponds to
particle-like pairing, with a monotonically increasing energy and a
positive effective pair mass, $B>0$ and $t_\text{B}>0$, so that $q=0$
is the bottom of the pair energy. For the $a_0<0$ case
(Fig.~\ref{fig:Fig0}(c)), this dispersion flips upside down into the
blue-dashed hole mode. This corresponds to hole-like pairing, for
which $q=0$ becomes a local maximum, with $B<0$ and $t_\text{B}<0$,
similar to the hole band in a semiconductor.  In case of a pure
lattice, one could flip the sign of $a_0$ via a particle-hole
transformation so that this blue-dashed line is flipped back to become
positive as the dispersion for hole pairs. However, for our present
case, due to the presence of the continuum dimension, there is no easy
way to do a particle-hole transformation so that we have to stay with
the (black solid) gapped positive branch, which is a flip of the hole
branch in Fig.~\ref{fig:Fig0}(a), as the dispersion of particle-like
Cooper pairs. When $a_0=0$, the two branches become symmetric, without
a gap.  For all three cases, the coefficients of the $q^2$ terms in
the inverse $T$ matrix expansion, $\xi^2$ and $\xi_{xy}^2$, must be
positive. (Note that $a_1$ is always positive.) Indeed, as shown in
Figs.~\ref{fig:Fig0}(d-f), for a negative $\xi^2$, the dispersion
$\tilde{\Omega}_\mathbf{q}$ of both particle-like
(Figs.~\ref{fig:Fig0}(d)) and hole-like (Figs.~\ref{fig:Fig0}(f))
pairs quickly become diffusive and thus cease to exist, unless higher
order terms, e.g., the $q^4$ terms, are included. In that case, the
pair dispersion will reach a minimum at a non-zero $q$. Our numerics
shows that in 2DOL, $\xi^2$ in the continuum dimension remains positive in
general but $\xi_{xy}^2\propto a_0t_\text{B}$ in the lattice plane may
indeed change sign so that $\xi_{xy}^2>0$ will constitute another stability
requirement for the superfluid phase.

Finally, the superfluid density must also be positive definite in a
stable superfluid. This, however,  has been found to be a relatively weaker
constraint in the cases of 3D continuum \cite{chien2006PRL,chen2006PRA}.

\subsection{Superfluid density}

As a representative transport property, superfluid density is an
important quantity in the superfluid phase. While it is always given
by $n/m$ at zero $T$ for the balanced case in 3D continuum, it will
take the average of the inverse band mass in the presence of a
lattice. Furthermore, in the presence of population imbalance, it may
become negative \cite{chien2006PRL,chen2006PRA,he2007PRB}, signaling
an instability of the superfluid state.  Here we shall also
investigate the behavior of the anisotropic superfluid density
$(n_\text{s}/m)$, and pay close attention to the population imbalanced case
and the situations where it  becomes negative.

The expression for superfluid density can be derived using the
linear response theory.  Following
Refs.~\cite{chien2006PRL,chen2006PRA,chen1998PRL,he2007PRB,ChenPhD},
we obtain for zero $T$ 
\begin{equation}
  \label{equ:pSF}
  \left(\frac{n_\text{s}}{m}\right)_{i}=\sum_\mathbf{k}\frac{\Delta^{2}_\text{sc}}{E_\mathbf{k}^2}\left[\frac{\Theta(E_{\textbf{k}\uparrow})}{E_\mathbf{k}}
  -\delta(E_{\textbf{k}\uparrow})\right]
  \left(\frac{\partial \xi_\mathbf{k}}{\partial{k}_i}\right)^2\,,
\end{equation}
where $i=x,y$ and $z$ for the lattice and the continuum directions,
respectively.
%$(\partial\xi_\mathbf{k}/\partial{k}_x)^2=4t^2d^2\sin^2(k_xd)$ and
%$(\partial\xi_\mathbf{k}/\partial{k}_y)^2=4t^2d^2\sin^2(k_yd)$, thus
%$(n_\text{s}/m)_x = (n_\text{s}/m)_y$.  For the momentum in the $z$ direction,
%$(\partial\xi_\mathbf{k}/\partial{k}_z)^2=k_z^2/m^2$.

%Numerical results and discussions
\section{Numerical results and discussions}

Due to the multiple tunable parameters for the present 2DOL,
the compete multidimensional phase diagram can be extremely complex.
Therefore, we shall focus on the lattice effect for the $p=0$ case,
together with the population imbalance for the $p \neq 0$ case, to give
several representative and informative phase diagrams. For our
numerics, it is convenient to define Fermi momentum
$k_\text{F}=(3\pi^{2}n)^{1/3}$ and Fermi energy
$E_\text{F}\equiv k_{B}T_\text{F}=\hbar^{2}k_\text{F}^{2}/2m$, as the units of
momentum and energy, respectively, which also sets $2m=1$. Note,
however, that this $E_\text{F}$ is \emph{not} equal to the chemical potential in
the noninteracting limit.

\subsection{Fermi surfaces in the noninteracting limit}

Fermi surface plays an important role in the superfluid and pairing
behavior of atomic Fermi gases. For 2DOL, it is very different from
the 3D continuum or 3D lattice case, so is it from 1DOL
\cite{zhang2017SR,wang2020PRA,wang2020PRA2}. This will lead to
different physics. Here we first present the shape and topology of the
Fermi surface for a series of representative sets of lattice
parameters ($t,d$). Shown in Fig.~\ref{fig:Fig1} is the
typical evolution behavior of the Fermi surface, calculated
self-consistently in the noninteracting limit at zero temperature. The
top row shows the evolution with the lattice constant, for
$k_\text{F}d=1, 2, 3$ and $4$ at fixed hopping integral $t/E_\text{F}=0.05$. Then
the bottom row shows the effect of hopping integral, with
$t/E_\text{F}=0.01$, 0.04, 0.07, and $0.1$ and fixed $k_\text{F}d=3$.

The lattice constant $d$ provides a confinement in the momentum space;
the larger $d$ the stronger confinement.  The top row in
Fig.~\ref{fig:Fig1} suggests that the Fermi surface becomes thicker
along the $z$ direction as $d$ increases for fixed $t$. Indeed,
fermions feel a stronger confinement in the lattice dimensions with a
shrinking first Brillouin zone (BZ), as $k_\text{F}d$ increases from 1 to 4,
and thus need to occupy higher $k_z$ states to keep the Fermi volume
unchanged, so that the noninteracting fermionic chemical potential is
pushed up. As a rough estimate, the maximum occupied $k_z$ increases
by a factor of 16 from left to right.  For relatively small $t/E_\text{F}=0.05$, the shape and topology of the Fermi surface
evolve from a closed plate for $k_\text{F}d = 1$ into one with only the top
and bottom faces while completely open on the four sides at the
BZ boundary of the lattice dimensions for $k_\text{F}d = 3$ and 4. For the
intermediate $k_\text{F}d=2$, the Fermi surface is open only at the center of
the four side faces at the BZ boundary. At the same time, the
effective filling factor in the lattice dimensions increases to
nearly unity as $k_\text{F}d$ increases from 1 to 4. In this way, for large
$d$, fermion dispersion on the Fermi surface on average becomes
hole-like in the lattice plane, while it always remains particle-like 
in the continuum dimension.

On the other hand, a smaller $t$ makes the fermion energy less
dispersive in the lattice dimensions, and thus the lattice band
becomes narrower and more fully filled.  In other words,  fermions
will tend not to go to higher $k_z$ states until the BZ at lower $k_z$
is fully occupied, leading to a flatter top and bottom of the
Fermi surface. This will also pull down the noninteracting fermionic
chemical potential.
As shown in the bottom row in Fig.~\ref{fig:Fig1}, the Fermi surface
becomes thinner and flatter in the $z$ direction as $t/E_\text{F}$ decreases
from $0.1$ to $0.01$ for fixed $k_\text{F}d = 3$.  In contrast,
the $t/E_\text{F}=0.07$ and $0.1$ cases have a much more dispersive
Fermi surface as a function of the in-plane momentum $(k_x,k_y)$.
Fermions at high $(k_x,k_y)$ states are removed for relatively large
hopping integral $t/E_\text{F}=0.07$ and $0.1$.

The evolution of the Fermi surface reveals that the in-plane fermion
motion on the Fermi surface becomes hole-like for relatively small $t$
and large $d$.  As a result, the nature of the in-plane and overall
pairing in this case will also change from particle-like to hole-like
when the contributions from lattice dimensions are dominant in the BCS
and unitary regimes \cite{sun2021AdP}.

It should be mentioned that in the strong pairing regime, the detailed
shape of the Fermi surface is no longer relevant, as pairing extends
essentially to the entire momentum space. However, the confinement in
the momentum space imposed by the lattice periodicity is always
present and will govern the physical behavior in the BEC regime.

\subsection{Phase diagram for the population balanced case}

\begin{figure}
  \includegraphics[clip,width=3.2in]{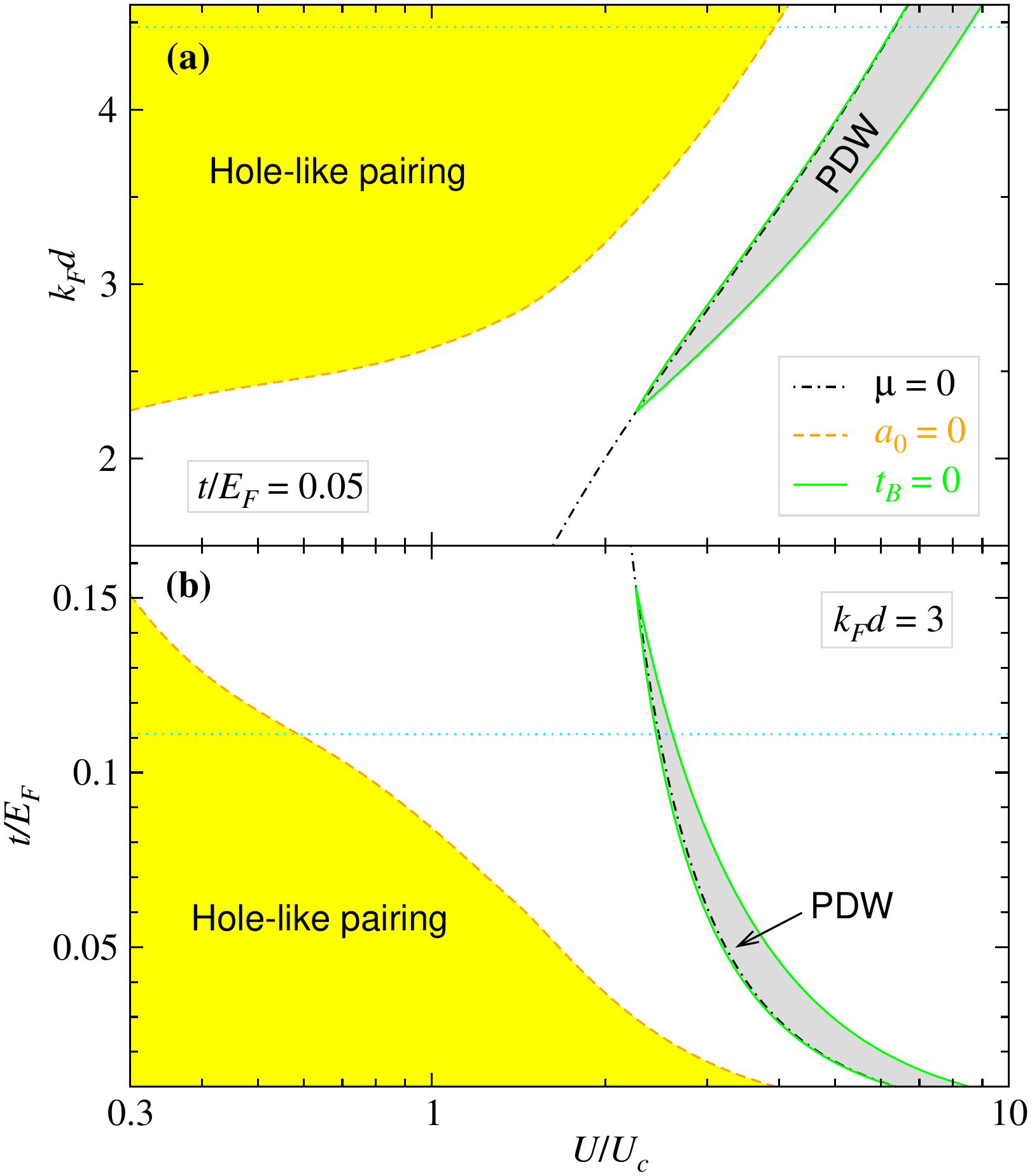}
  \caption{ Phase diagram in the balanced case in the (a) $d$ -- $U$
    plane for $t/E_\text{F}=0.05$ and (b) in the $t$ -- $U$ plane for
    $k_\text{F}d=3$.  The (orange dashed) $a_0=0$ curve separates hole-like
    pairing (yellow shaded region) on the left from particle-like
    pairing on the right. Enclosed inside the (green) $t_\text{B}=0$ line is
    a PDW ground state (grey shaded region).  Also plotted is the
    (black dot-dashed) $\mu=0$ line.  The (cyan) dotted line denotes
    the upper limit for (a) $d$ and (b) $t$, respectively, as defined
    by $2mtd^2\leq 1$. }
  \label{fig:Fig2}
\end{figure}

It is known from the 3D continuum case that the balanced case and the
imbalanced case with $p \rightarrow 0^+$ are not continuously
connected in the BCS and unitary regimes at $T=0$
\cite{chien2006PRL,chen2007PRB}. Population imbalance leads to very
distinct behaviors. Therefore, we present in this section the balanced
results only.

In Fig.~\ref{fig:Fig2}, we present a typical phase diagram (a) in the
$d$ -- $U$ plane, for fixed relatively small $t/E_\text{F}=0.05$, and
(b) in the $t$ -- $U$ plane, for relatively large $k_\text{F}d=3$,
corresponding to the cases of the top and bottom rows in
Fig.~\ref{fig:Fig1}, respectively.  The lattice constant in panel (a)
ranges from relatively small $k_\text{F}d=1$ with $2mtd^2=0.05$ to the
upper limit $k_\text{F}d=2\sqrt{5}$ with $2mtd^2=1$ denoted by the
horizontal (cyan) dotted line, and the hopping integral in panel (b)
ranges from relatively small $t/E_\text{F}=0.01$ with $2mtd^2=0.09$ to
the upper limit $t/E_\text{F}=1/9$ with $2mtd^2=1$ denoted by the
horizontal (cyan) dotted line.  In either panel, the (black
dot-dashed) $\mu=0$ curve defines the boundary between the fermionic
and the bosonic regimes.  The (yellow) shaded region on the left of
the (orange) dashed $a_0=0$ curve is a hole-like pairing regime with
$a_0<0$, whereas the overall pairing evolves from hole-like into
particle-like with $a_0>0$ across the $a_0=0$ curve.  A PDW ground
state with $t_\text{B} < 0$ emerges within the grey shaded region,
enclosed within the (green) $t_\text{B} = 0$ curve. The entire phase
space is a superfluid except for the PDW phase. Note that the PDW phase
usually starts immediately before $\mu$ decreases down to zero, as the
pairing strength increases. The fact that there are two branches of
the $t_\text{B}=0$ curve indicates that there is an reentrant behavior
of $T_\text{c}$ as a function of pairing strength. In the absence of
population imbalance, similar reentrant behavior of superfluidity and
associated PDW ground state have \emph{not} been found in any other
balanced systems with a short range pairing interaction, except in a
very narrow range of density slightly above 0.53 in the attractive
Hubbard model
\cite{ChenPhD,PhysRevA.77.011601,PhysRevA.78.043612}. With a
long-range anisotropic dipole-dipole interaction, however, such a
reentrant behavior and PDW state have been predicted in the $p$-wave
superfluid in dipolar Fermi gases \cite{che2016PRA}.

As shown in Fig.~\ref{fig:Fig2}, the interaction range for hole-like
pairing extends toward stronger pairing regime with (a) increasing $d$
or (b) decreasing $t$. This can be explained by the evolution of the
shape and topology of the Fermi surface, as shown in
Fig.~\ref{fig:Fig1}.  As $d$ increases or $t$ decreases, the Fermi
surface gradually opens up at the four $X$ or $Y$ points located at
$(k_x,k_y) = (\pm\pi/d,0)$ and $(0, \pm\pi/d)$, and becomes fully open
at the first BZ boundary for large $d$ small $t$, leading to an
effective filling factor above $1/2$ in the lattice dimensions. In
contrast to the 1DOL case, the existence of two lattice dimensions is
enough to dominate the contributions of the remaining one continuum
dimension (which is always particle-like due to its parabolic fermion
dispersion), so that both the in-plane and the overall pairing becomes
hole-like when $d$ is large or $t$ is small, with $a_0<0$ in the
linear frequency term of the inverse $T$ matrix expansion. This is
especially true in the weak coupling regime, where the superfluidity
is more sensitive to the underlying Fermi surface. As the interaction
goes stronger toward the BEC regime, the gap becomes large and the
Fermi level (i.e., chemical potential $\mu$) decreases and then becomes
negative, hence the shape of the non-interacting Fermi surface is no
longer important. In this case, the contributions from the lattice
dimensions will spread evenly across the entire BZ, so that the
continuum dimension will become dominant, and the overall pairing
eventually changes from hole-like to particle-like (with $a_0 > 0$).
As shown in Fig.~\ref{fig:Fig1}, within the occupied range of $k_z$,
the average (or effective) filling factor within the first BZ in the
$xy$ plane increases with increasing $d$ and/or decreasing
$t$. Therefore, as $d$ increases, or $t$ decreases, the effect of the
above-half-filling status persists into stronger pairing regime, and
thus the hole-like pairing region in Fig.~\ref{fig:Fig2}
extends toward right.

\begin{figure}
  \includegraphics[clip,width=3.1in]{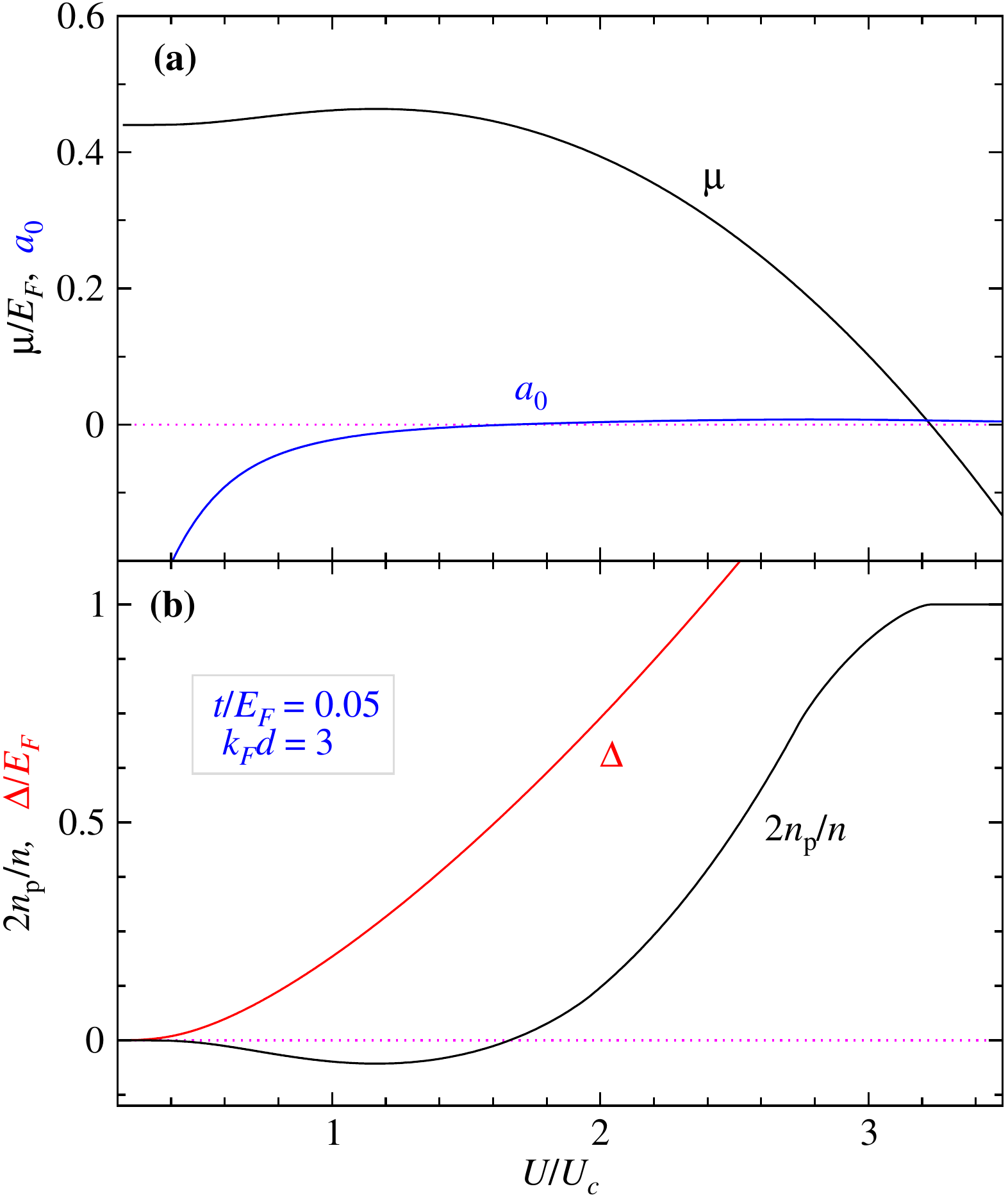}
  \caption{Behaviors of (a) $\mu$ and $a_0$  and (b) $2n_\text{p}/n$ and $\Delta$ as a function of $U/U_\text{c}$ for  $t/E_\text{F}=0.05$ and  $k_\text{F}d=3$ without population imbalance. The maximum of $\mu$ corresponds to the minimum of $2n_\text{p}/n$.}
  \label{fig:Fig2.5}
\end{figure}

Shown in Fig.~\ref{fig:Fig2.5} is the behavior of (a) $\mu$ as a
function of $U$, along with (b)  $2n_\text{p}/n$, where
$n_\text{p}\equiv a_0\Delta^2$, for $t/E_\text{F}=0.05$ and $k_\text{F}d=3$. Also plotted
are $a_0$ and $\Delta$.  This corresponds to a horizontal cut at
$k_\text{F}d=3$ in Fig.~\ref{fig:Fig2}(a) or at $t/E_\text{F}=0.05$ in
Fig.~\ref{fig:Fig2}(b).  Inside the hole-like pairing regime, $a_0<0$
and thus the chemical potential $\mu$ goes above its noninteracting
value.  This can be seen from the expression \cite{ChenPhD,sun2021AdP}
\begin{equation}
  \sum_\mathbf{k} \Theta(-\xi_\mathbf{k})  =  n/2 - a_0\Delta^2\,.
  \label{eq:np}
\end{equation}
The chemical potential $\mu$ increases with the pairing strength,
until it reaches a maximum where $n_\text{p} $
reaches a minimum. Here $2n_\text{p}/n$ is roughly the pair fraction,
which reaches unity in the BEC regime. This plot is very close to its
counterpart at $T_\text{c}$, which can be found in
Ref.~\cite{sun2021AdP}, since the temperature dependencies of both
$\mu$ and $a_0$ are weak, except that here $a_0$ changes sign at a slightly
larger $U/U_\text{c}$. As usual, the excitation gap $\Delta$ increases
with $U/U_\text{c}$.

The PDW ground state in Fig.~\ref{fig:Fig2} with $t_\text{B} < 0$ at a
intermediate coupling strength for (a) relatively large $k_\text{F}d$
with fixed $t/E_\text{F}=0.05$ or (b) small $t$ with fixed
$k_\text{F}d=3$ is associated with the strong inter-pair repulsive
interaction, relatively large pair size and high pair density.  Close
to $\mu = 0$, nearly all fermions have paired up with a relatively
large pair size and a heavy effective pair mass, and the inter-pair
repulsive interaction becomes strong. A large $d$ or small $t$
strongly suppresses the pairing hopping kinetic energy, and the large
pair size and high pair density strongly reduce the pair mobility. All
these factors lead to Wigner crystallization and hence PDW in the $xy$
plane, which can also be called a Cooper pair insulator.  The negative
sign of $t_\text{B}$ within the grey shaded region indicates that the
minimum of the pair dispersion $\tilde{\Omega}_{\textbf{q}}$ has
shifted from $\mathbf{q}=0$ to $\mathbf{q} = (\pi/d,\pi/d,0)$, with
crystallization wave vector $(q_x, q_y)$ in the $xy$ plane.  As the
pairing interaction increases in the BEC regimes, the pair size
shrinks and inter-pair repulsive interaction becomes weak; hence
$t_\text{B}$ changes from negative back to positive, corresponding to a
quantum phase transition from a PDW insulator to a superfluid.

\begin{figure}
  \includegraphics[clip,width=3.1in]{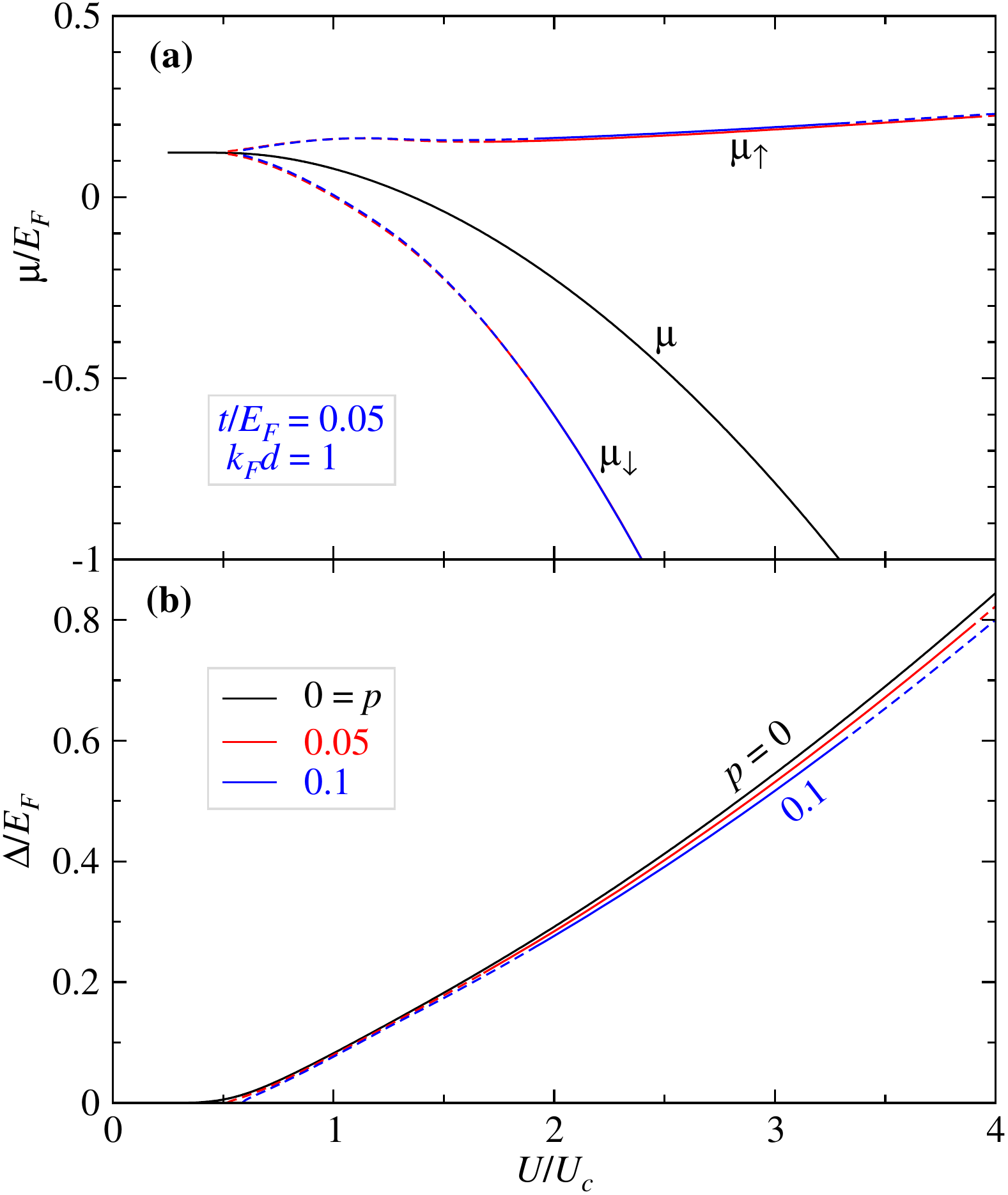}
  \caption{Behaviors of (a) $\mu$ or $\mu_\sigma$ and (b) $\Delta$ as a function of $U/U_\text{c}$ for $p=0$ (black), 0.05 (red) and 0.1 (blue lines), with fixed $t/E_\text{F}=0.05$ and  $k_\text{F}d=1$. Here solid and dashed lines denote stable and unstable solutions, respectively.}
  \label{fig:Fig3}
\end{figure}

\begin{figure}
  \includegraphics[clip,width=3.2in]{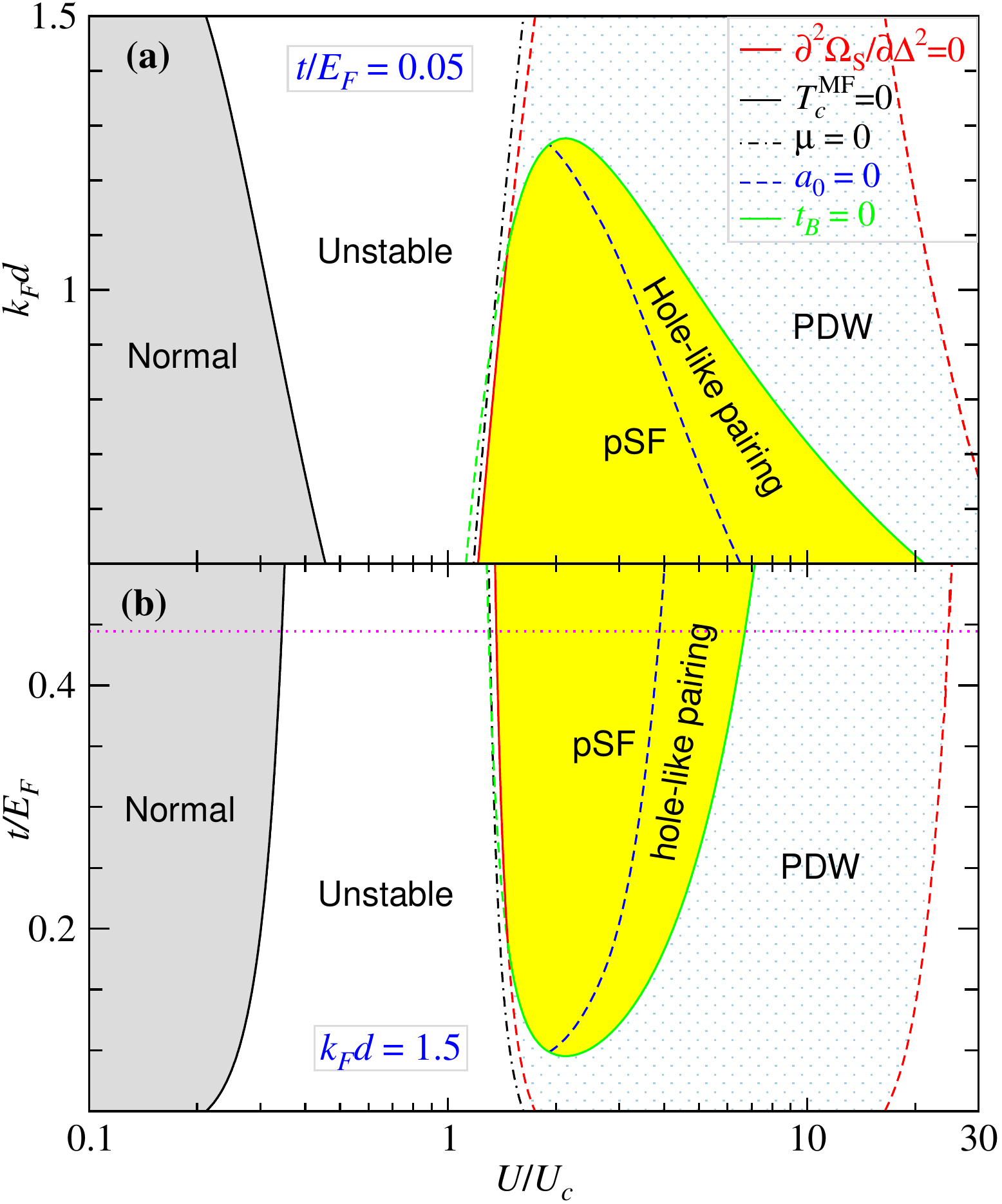}
  \caption{ Phase diagrams at $p=0.001$ in the (a) $d$ -- $U$ plane with $t/E_\text{F}=0.05$ and in the (b) $t$ -- $U$ plane with $k_\text{F}d=1.5$, respectively. As labeled, the solid lines along with the (red) stability line split the diagram into four phases: Normal gas (grey shaded, on the left of the black $T_\text{c}^\text{MF}=0$ line), unstable mean-field superfluid (unshaded), PDW phase (dot shaded), and stable polarized superfluid (yellow shaded region, bounded by the green $t_\text{B}=0$ line). Pairing on the right of the $a_0=0$ line (blue dashed)  has a hole-like nature. The chemical potential $\mu=0$ line (black dot-dashed) separates the fermionic regime (one the left) from the bosonic regime (on the right). The (magenta) dotted line sets the upper bound for $t$ via $2mtd^2\leq 1$.}
  \label{fig:Fig4}
\end{figure}

Combining Figs.~\ref{fig:Fig1} and \ref{fig:Fig2}, we find that the
emergence of hole-like pairing and the PDW phase is associated with
the open Fermi surface topology. Once the Fermi surface is closed,
both hole-like pairing and the PDW phase disappear.

In case of a closed Fermi surface, typical behaviors of the chemical
potential $\mu$ and the excitation gap $\Delta$ for the balanced case
can be seen from the $p=0$ lines in Fig.~\ref{fig:Fig3}, calculated
for $t/E_\text{F}=0.05$ and $k_\text{F}d=1$. Here $\mu$ decreases monotonically with
$U/U_\text{c}$. Without a hole-like pairing regime, these solutions look
qualitatively similar to other cases, e.g., in 3D continuum or 3D
lattice, except that they follow a different asymptotic behavior in
the BEC limit \cite{sun2021AdP}.

\subsection{Phase diagram for the population imbalanced case}

We now proceed and present our results for the population imbalanced
case. With the added parameter $p$, the phase diagram becomes much
more complicated. It renders the otherwise superfluid state unstable
in the vast areas in the phase space.

To make the comparison easier, we begin by presenting phase diagrams
in Fig.~\ref{fig:Fig4} in the same (a) $d$ -- $U$ and $t$ -- $U$
planes, as in Fig.~\ref{fig:Fig2}, but with a tiny nonzero $p =
0.001$. Here a normal gas phase (grey shaded) emerges in the weak
coupling regime, delineated by the (black solid) $T_\text{c}^\text{MF}=0$ line,
which is given by Eqs. (\ref{eq:pgap})-(\ref{eq:neqb}) with
$\Delta = 0$. Indeed, in the presence of an imbalance, pairing cannot
take place for an arbitrarily weak interaction.  There exists a stable
pSF phase (yellow shaded), defined by the (green solid) $t_\text{B}=0$ line
and further confined by the stability condition (red solid line). The
pSF phase resides in the low $d$ and large $t$ regime. A PDW ground
state emerges in the dot shaded region, enclosed by the $t_\text{B}=0$ line
and the dashed part of the (red) stability line. Then the rest
unshaded space allows for an unstable mean-field superfluid solution,
which may yield to phase separation. Now that the underlying lattice
in the $xy$ plane breaks the continuous translational symmetry, the
exotic Fulde-Ferrell-Larkin-Ovchinnikov (FFLO) states may possibly
exist in part of the unstable region
\cite{FF,*LO,WangLOFF,Samokhin2006PRB}.

One can immediately tell that the vertical axes in Fig.~\ref{fig:Fig4}
take different parameter ranges from those in Fig.~\ref{fig:Fig2},
even though the imbalance $p=0.001$ is very small. While the $d$ --
$U$ phase diagram in Fig.~\ref{fig:Fig4}(a) is still calculated with
$t/E_\text{F}=0.05$, the stable pSF phase is now restricted to relatively
small $d$ (yellow shaded area). However, the $t$ -- $U$ phase diagram
has to be calculated at a much smaller $d$, with $k_\text{F}d = 1.5$, as
there is no stable pSF phase for $k_\text{F}d=3$ within the constraint
$2mtd^2\leq 1$ (i.e., $t/E_\text{F}\le 1/9$). In both cases in
Fig.~\ref{fig:Fig4}, the Fermi surface is closed. Unlike the balanced
cases, one cannot find a stable superfluid solution with an open Fermi
surface. For this reason, one does not find a hole-like pairing region
in the weak coupling regime, but rather one in the strong coupling
regime, on the right of the (blue dashed) $a_0=0$ line. Note that in
the superfluid phase of hole-like pairing (on the right of the blue
dashed line), both $a_0$ and $t_\text{B}$ are negative but the product
$\xi_{xy}^2$ is positive. Outside the $t_\text{B}=0$ curve, we
have $\xi_{xy}^2<0$, so that the mean-field superfluid solution
becomes unstable, yielding to the PDW phase.
% , phase separation or FFLO.
The smallness of $p$ suggests that the ground state of
$p \rightarrow 0^+$ is \emph{not} continuously connected to the $p=0$
case, consistent with that in 3D continuum \cite{chien2006PRL}. In
comparison with Fig.~\ref{fig:Fig2}, the current large PDW phase in the bosonic regime is
totally a consequence of population imbalance.

\begin{figure}
  \includegraphics[clip,width=3.2in]{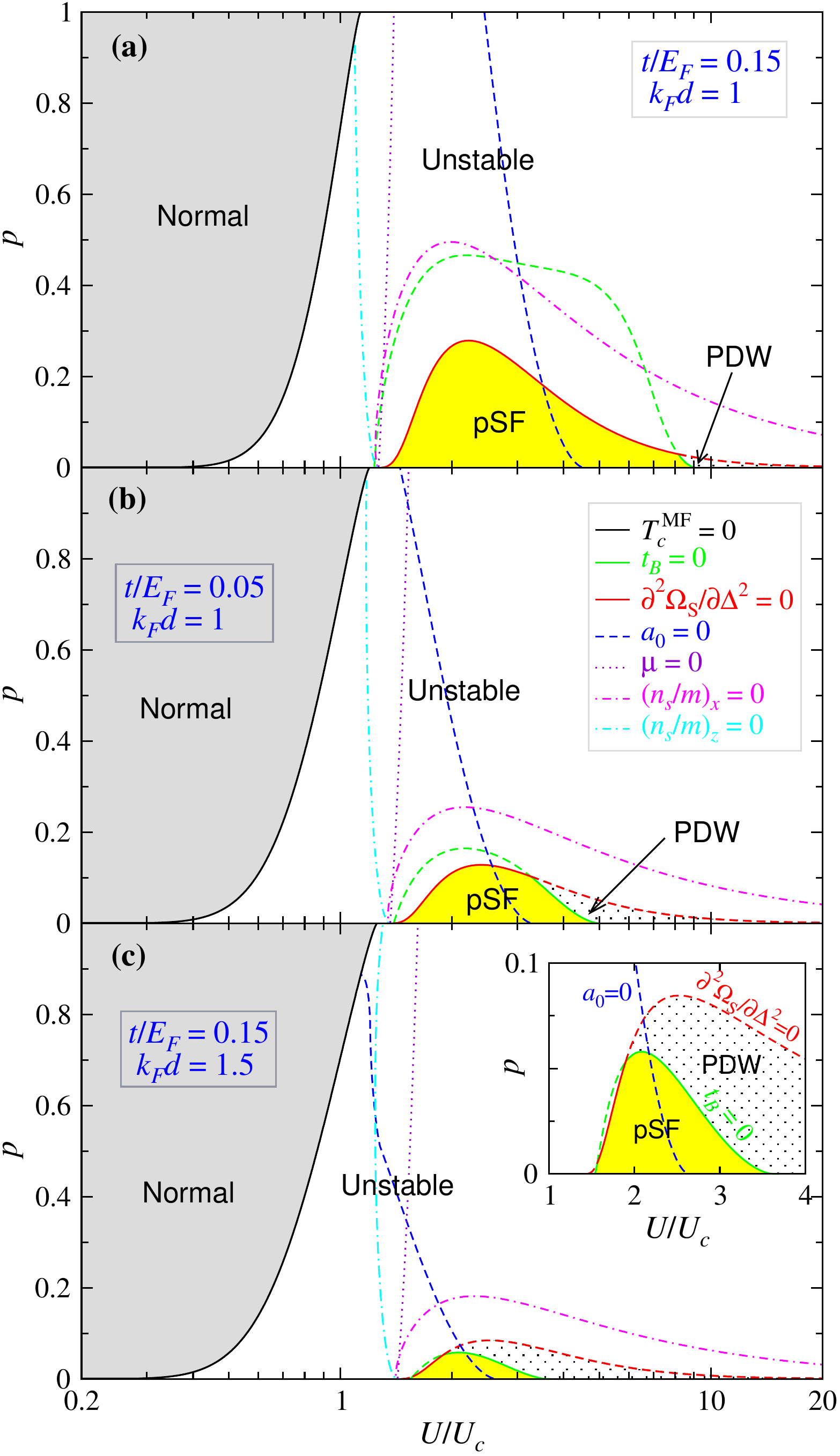}
  \caption{Phase diagrams in the $p$ -- $U$ plane for (a)
    $(t/E_\text{F}, k_\text{F}d) =(0.15, 1)$, (b) (0.05, 1), and (c) (0.15,
    1.5). The solid $T_\text{c}^\text{MF}=0$ (black) and $t_\text{B}=0$ (green)
    lines, as well as the (red) stability
    $\partial^2\Omega_\text{S}/\partial\Delta^2=0$ line (both solid
    and dashed) divide the plane into four phases: Normal gas (grey
    shaded), unstable superfluid (unshaded), PDW phase (dot shaded),
    and stable pSF phase. Across the $a_0=0$ line (blue dashed) the
    pairing nature changes from particle-like (on the left) to
    hole-like (on the right). The $\mu=0$ line (violet dotted)
    separate fermionic (left) from bosonic (right) regimes. Also
    plotted are lines of the superfluid density $(n_\text{s}/m)_x=0$ (magenta
    dot-dashed) in the $x$ direction, and $(n_\text{s}/m)_z=0$ (cyan
    dot-dashed) in the $z$ direction. Superfluid density is negative
    on the weaker coupling or larger $p$ side of these curves. Shown
    in the inset of panel (c) is a zoom-in of the pSF phase.}
  \label{fig:Fig5}
\end{figure}

Now we take $p$ as a varying parameter and explore phase diagrams in
the $p$ -- $U$ plane. 
Shown in Fig.~\ref{fig:Fig5} are the phase diagrams for (a)
$(t/E_\text{F}, k_\text{F}d) =(0.15, 1)$, (b) (0.05, 1), and (c) (0.15, 1.5). 
Panels (b) and (c) show the effect of changing $t$ and $d$,
respectively. In all three cases, there are three different phases,
delineated by solid lines, as well as a PDW phase. A normal gas phase
(grey shaded) takes the weaker coupling and larger $p$ area, on the
left of the $T_\text{c}^\text{MF} = 0$ curve. The vast majority is an
unstable mean-field superfluid (unshaded), which should yield to phase
separation or FFLO solutions. The stable pSF phase (yellow shaded)
occupies only a small area. Finally, the PDW phase (dot shaded) takes
the small region next to the pSF phase, bounded by the (red dashed)
stability $\partial^2 \Omega_\text{S}/\partial\Delta^2 = 0$ line and
(green sold) $t_\text{B}=0$ line. When compared with panel (a), one readily
sees that the pSF phase shrinks as $t$ decreases (panel (b)) and/or as
$d$ increases (panel (c)). This is because both increasing $d$ and
reducing $t$ lead to stronger momentum confinement in the lattice
dimensions. In agreement with Fig.~\ref{fig:Fig4}, the Fermi surface
for all these three cases are closed. Note that the (red) stability
line and the (green) $t_\text{B}=0$ line cross into each other, and the pSF
phase is bounded by the stronger of these two conditions. Here also
plotted are the lines along which the superfluid density vanishes. As
found in 3D continuum, the positivity of superfluid density
constitutes a much weaker stability constraint, as both lines of
$(n_\text{s}/m)_x=0$ in the lattice dimension and of $(n_\text{s}/m)_z=0$ in the
continuum dimension lie completely within the unstable area. Note that
while the $(n_\text{s}/m)_z=0$ line looks very similar to its 3D continuum
counterpart, the $(n_\text{s}/m)_x=0$ line exhibits an unusual nonmonotonic
behavior,  caused by the lattice effect. From the (violet dotted) $\mu=0$
curve, one readily sees that, as in Fig.~\ref{fig:Fig4}, the pSF phase resides
completely within the bosonic regime.

The fact that the pSF phase exists only in a small bosonic region (in
both Fig.~\ref{fig:Fig4} and Fig.~\ref{fig:Fig5}) is in stark contrast
with the 3D continuum case, for which the stability line
$\partial^2 \Omega_\text{S}/\partial\Delta^2 = 0$ extends
monotonically up to $p=1$, and a polarized superfluid exists for
arbitrary imbalance $p$ in the BEC regime
\cite{chien2006PRL}. Apparently, this difference can be attributed to
the presence of two lattice dimensions. Indeed, for 1DOL, with only
one lattice dimension, the stability line already cannot extend to
$p=1$. However, the pSF phase in 1DOL can extend all the way to the
deep BEC limit \cite{Chen2020cpl}. This is also supported by the fact
that with three lattice dimensions in a 3D attractive Hubbard model,
one can barely find a pSF state except at very low density and
extremely low $p$ \cite{Micnas14AP}. Therefore, one can conclude that
more lattice dimensions make it more difficult to have a stable
pSF ground state.

\begin{figure*}
  \includegraphics[clip,width=6in]{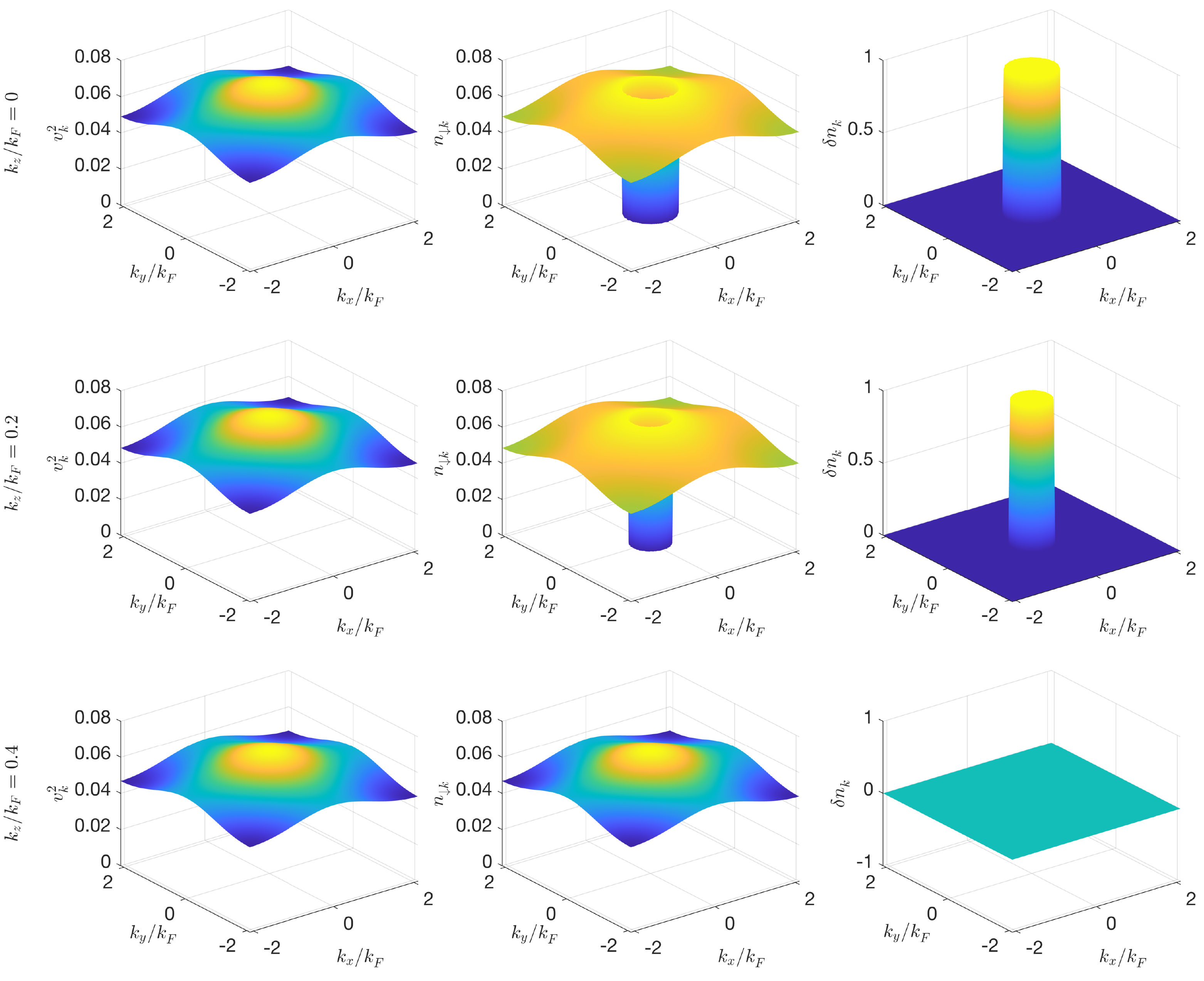}
   \caption{Momentum distributions of $v_\mathbf{k}^2$ (left),
     $n_{\mathbf{k}\downarrow}$ (middle) and $\delta n_\mathbf{k}$
     (right column) in the $(k_x,k_y)$ plane at different
     $k_z/k_\text{F} = 0$ (top), 0.2 (middle) and 0.4 (bottom row), with
     $U/U_\text{c}=4$ and $p=0.05$, for $t/E_\text{F}=0.15$ and $k_\text{F}d=1.5$. The
     excessive fermion distribution, $\delta n_\mathbf{k} $, occupies
     the low in-plane momentum part and below $k_z/k_\text{F} = 0.4$ (right
     column), $v_\mathbf{k}^2$ (left column) remains roughly constant
     in the entire BZ and for $|k_z/k_\text{F}| \leq 0.4$, and
     $n_{\mathbf{k}\downarrow}$ (middle column) is given by
     $v_\mathbf{k}^2$ but with the central part expelled.  }
  \label{fig:nk}
\end{figure*}

This phenomena can be easily understood from the momentum distribution
of paired fermions, which would be given by $v_\mathbf{k}^2$ had there
been no imbalance. In 3D continuum, $v_\mathbf{k}^2$ in the deep BEC
regime extends to the entire infinitely large momentum space in all
directions, leading to a vanishingly small occupation for paired
fermions. Therefore, the excessive majority fermions can readily
occupy the low momentum states, with essentially no Pauli blocking
from paired fermions. However, when one or more lattice dimensions are
present, the momentum in these dimensions is restricted to the first
BZ, so that $v_\mathbf{k}^2$ in these dimensions cannot be infinitesimally
small even in the extreme BEC limit, which will cause a repulsion to
excessive majority fermions. This repulsion increases with $p$, and
may become costly enough so as to render the mean-field superfluid
solution unstable.  As a result, the
distribution of paired fermions is now roughly given by that of the minority fermions,
$n_{\mathbf{k}\downarrow} =
\Theta(E_{\mathbf{k}\uparrow})v_\mathbf{k}^2$, which reduces to
$v_\mathbf{k}^2$ for $p=0$.
%Therefore, in the presence of imbalance,
%$v_\mathbf{k}^2$, defined via the spin-averaged dispersion
%$\xi_\mathbf{k}$ and $E_\mathbf{k}$, no longer reflects the momentum
%distribution of paired fermions.

Unlike the $p=0$ case, for which hole-like pairing takes place in the
weaker coupling regime when $t$ is small and/or $d$ is large, here
hole-like pairing occurs in the BEC regime via a completely different
mechanism. As mentioned above, all the three cases shown in
Fig.~\ref{fig:Fig5} have a closed noninteracting Fermi surface. As the
pairing becomes stronger, the momentum distribution of $v_\mathbf{k}^2$ 
in the $xy$ plane extends to the entire first BZ, and becomes roughly
a constant at strong coupling; in the absence of population imbalance,
this would lead to a rough cancellation (via averaging over the inverse
fermion band mass) due to the particle-hole symmetry of the lattice
band. However, for any finite $p$, the excessive majority fermions
will tend to occupy the low $(k_x, k_y)$ states, and thus expel paired
fermions toward higher $(k_x, k_y)$ states, which have a negative
(i.e., hole-like) band mass, leading to a net hole-like contribution
to $a_0$ in the pair propagator, when integrated over the entire
BZ.  This also explains why the $a_0=0$ line leans
toward weaker coupling with increasing $p$.

Shown in Fig.~\ref{fig:nk} is an example of the momentum distributions
of $v_\mathbf{k}^2$ (left), $n_{\mathbf{k}\downarrow}$ (middle) and
$\delta n_\mathbf{k}$ (right column) in the $(k_x,k_y)$ plane at
different $k_z/k_\text{F} = 0$ (top), 0.2 (middle) and 0.4 (bottom row), with
$U/U_\text{c}=4$ and $p=0.05$, for $t/E_\text{F}=0.15$ and $k_\text{F}d=1.5$. This
corresponds to a PDW state in Fig.~\ref{fig:Fig5}(c). Indeed, the
excessive fermion distribution,
$\delta n_\mathbf{k} = \Theta(-E_{\textbf{k}\uparrow})$, occupies the
low in-plane momentum part and below $k_z/k_\text{F} = 0.4$ (right
column). In addition, $v_\mathbf{k}^2$ (left column) remains roughly 
constant in the entire BZ and for $|k_z/k_\text{F}| \leq 0.4$. Most
interestingly, the minority fermion distribution
$n_{\mathbf{k}\downarrow}$ (middle column) is given by
$v_\mathbf{k}^2$ but with a hole dug out at the center, due to the
Pauli repulsion with the excessive fermions.

As a representative example, we show in Fig.~\ref{fig:Fig3} the behavior of
(a) $\mu_\sigma$ and (b) the gap $\Delta$ for $p=0.05$ (red) and 0.1
(blue) with fixed $t/E_\text{F}=0.05$ and $k_\text{F}d=1$, as a function of $U$.
They correspond to horizontal cuts at $p=0.05$ and 0.1 in
Fig.~\ref{fig:Fig5}(b), and should be compared with the $p=0$ case
(black solid curves). The solid part of these lines are stable pSF
solutions, while the dashed lines are unstable mean-field
solutions. There are a few remarkable features. Firstly, the
excitation gap changes only slowly with imbalance $p$, except that it
does not have a solution below certain threshold of interaction
strength. Secondly, at given pairing strength, $\mu_\sigma$ for
$p=0.05$ and $p=0.1$ are very close to each other, but both far
separated from the $\mu$ curve for $p=0$. This again indicates that
the $p\rightarrow 0^+$ case is not continuously connected to the $p=0$
case; with a tiny bit of imbalance, $\mu_\uparrow$ and
$\mu_\downarrow$ immediately split up. Lastly, $\mu_\uparrow$
increases slowly with pairing strength in the BEC regime. This is
different from its counterpart in 3D continuum and 1DOL; for the
former, $\mu_\uparrow$ decreases while for the latter $\mu_\uparrow$
approaches a $p$-dependent constant asymptote, as the pairing strength
increases toward the BEC limit. This can be attributed to the
emergence of hole-like pairing (with $a_0<0$) in the strong pairing
regime as the number of lattice dimensions increases. To verify this
idea, we have also checked the mean-field solution for imbalanced
3DOL, and found that, indeed, $\mu_\uparrow$ also increases with the
pairing strength in the BEC regime at $T=0$, along with a
negative $a_0$.

\begin{figure}
  \includegraphics[clip,width=3.2in]{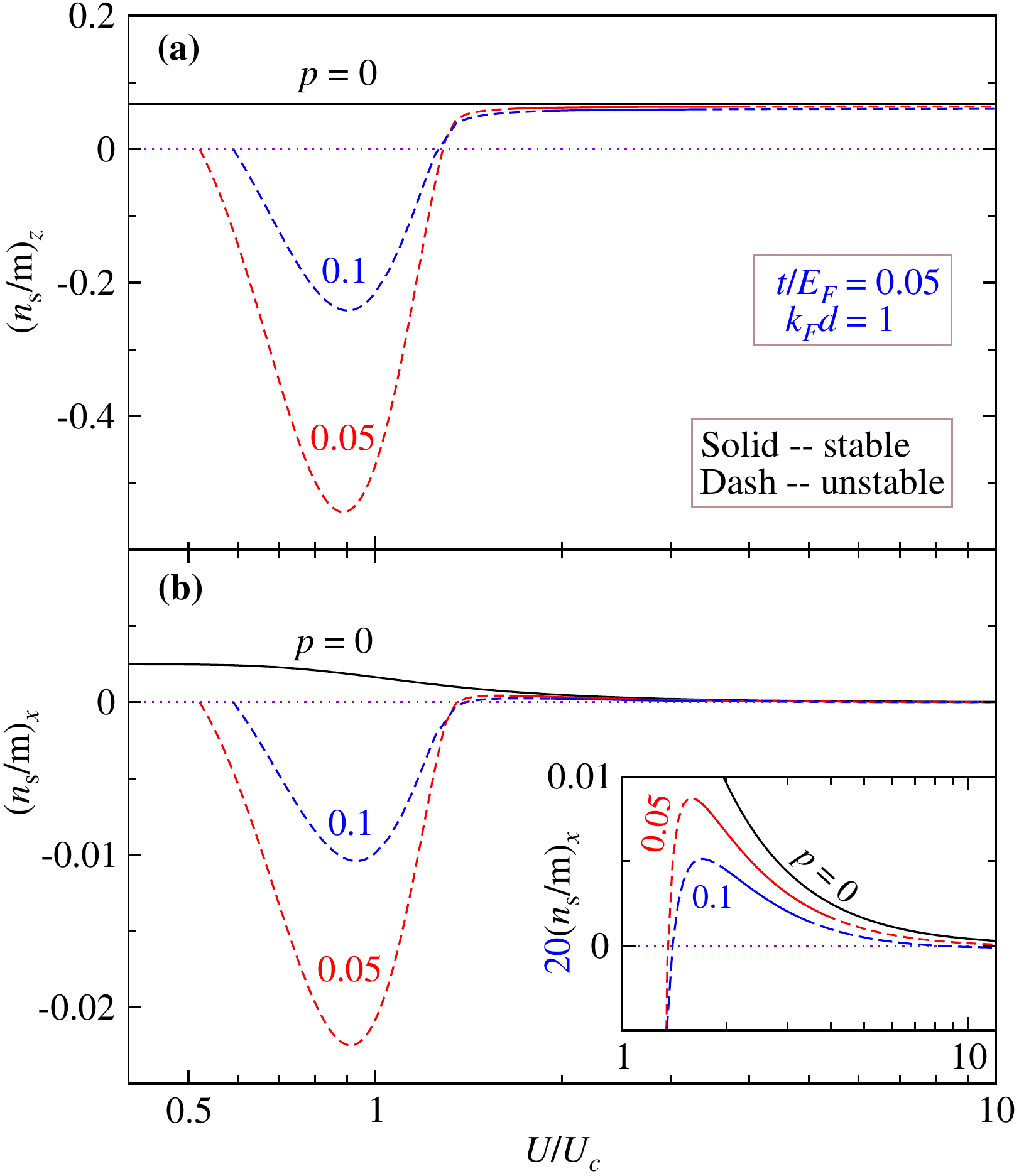}
  \caption{Superfluid density (a) $(n_\text{s}/m)_z$ and (b) $(n_\text{s}/m)_x$ as a function of $U/U_\text{c}$ for $p=0$ (black), 0.05 (red) and 0.1 (blue) at $t/E_\text{F}=0.05$ and $k_\text{F}d=1$. Solid (dashed) lines denote stable (unstable) solutions. Shown in the inset is 20 times magnified $(n_\text{s}/m)_x$ vs $U/U_\text{c}$. }
  \label{fig:Fig6}
\end{figure}

Finally, we present the typical behavior of the superfluid density in
the imbalanced case. Shown in Fig.~\ref{fig:Fig6} are (a) $(n_\text{s}/m)_z$
and (b) $(n_\text{s}/m)_x$ in the continuum and lattice dimensions,
respectively, as a function of $U/U_\text{c}$ for $p=0$, 0.05 and 0.1 at
fixed $t/E_\text{F}=0.05$ and $k_\text{F}d=1$. Here solid and dashed lines are
stable and unstable solutions, respectively. As expected, both are
always positive for the balanced case. In addition, $(n_\text{s}/m)_x$ is
much smaller than $(n_\text{s}/m)_z$, because it involves the average of the
inverse band mass. For the imbalanced case, the superfluid density 
deviates continuously from its positive $p=0$ value  as $p$ increases from 0.
However, in the unitary and weak coupling regimes, both continuum and
lattice components will become negative for $p\neq 0$. Furthermore,
the superfluid density is more negative for smaller (but finite)
$p$. This implies an immediate discontinuous jump from the $p=0$ value
to a large negative value for $p=0^+$ in this regime. Note that for
strong enough interaction, $(n_\text{s}/m)_x$ will again change sign to
negative, but gradually rather than abruptly, as can already be seen
from the $p=0.1$ curve. This has to do with the lattice induced
confinement in the momentum space and the Pauli exclusion between
paired and excessive fermions.

So far, it is not yet clear whether the PDW state can sustain a
superfluid order, with and without an imbalance. If the answer is yes,
then it will become a supersolid state rather than a Cooper pair
insulator. We leave this to a future study.

It should be noted that we have worked with a system with homogeneous 
fixed densities. For this reason, we have not chosen to use $\mu$ and
$h$ as control variables, which are more appropriate for systems
connected with a large reservoir so that the chemical potentials are
fixed or can be tuned separately. In such a case, all
$h< \sqrt{\min(0,\mu)^2+\Delta^2}$ corresponds to the population
balanced state. One can, however, convert between these two approaches,
by calculating corresponding densities (and Fermi energy) for given
$\mu$ and $h$, and performing a rescaling.

%Conculsions
\section{Conclusions}

In summary, we have studied the superfluid phase diagram of Fermi
gases with a short range pairing interaction in 2DOL at zero
temperature with and without population imbalance in the context of
BCS-BEC crossover.  We find that the mixing of lattice and continuum
dimensions, together with population imbalance, has an extraordinary
effect on pairing and the superfluidity of atomic Fermi gases. For the
balanced case, the ground state is a stable superfluid, except that
 a PDW ground state emerges for a finite range of intermediate
pairing strength in the case of relatively small $t$ and large $d$,
and the nature of the in-plane and overall pairing may change from
particle-like to hole-like in the BCS and unitary regimes for these
$t$ and $d$, which are associated with an open Fermi surface on the BZ
boundary of the lattice dimensions.  Thus the phase space for the PDW
ground state and hole-like pairing shrinks with increasing $t$ and/or
decreasing $d$.

For the imbalanced case, the presence of population imbalance has a
dramatic detrimental effect, in that the stable polarized superfluid
phase occupies only a small region in the bosonic regime in the
multi-dimensional phase space, and will shrink and disappear with
increasing $d$ and $p$ and decreasing $t$. The pSF phase can be found
only for relatively large $t$ and small $d$, associated with a closed
non-interacting Fermi surface, as well as for low $p$. In comparison
with 3D continuum, the presence of lattice dimensions introduces
confinement in the momentum space, which leads to strong Pauli
repulsion between paired and excessive fermions. Due to this
repulsion, the nature of pairing changes from particle-like to
hole-like in the strong pairing regime, and a PDW phase emerges next
to the pSF phase. In addition to the normal gas phase, stability
analysis shows that an unstable mean-field solution exists and may
yield to phase separation (and possibly FFLO) in the rest of the phase
diagram.
These findings for 2DOL are very different from pure 3D continuum,
3D lattices, and 1DOL, and should be tested in future experiment.
%can be a milestone in future theoretical and
%experimental studies.

%Ackownledgments
\section{Acknowledgments}
%\begin{acknowledgments}
This work was supported by the National Natural Science Foundation of
China (Grant No. 11774309), the Innovation Program
for Quantum Science and Technology (Grant No. 2021ZD0301904), as well
as the University of Science and Technology of China.
%\end{acknowledgments} 


\end{document}